\documentclass[11pt]{article}
\usepackage{fullpage}

\usepackage[english]{babel}

\usepackage{amssymb}
\usepackage{mathtools}
\usepackage{hyperref}

\newcommand{\E}{\mathbb{E}}
\newcommand{\Prob}{\mathbb{P}}
\newcommand{\indep}{{\perp\!\!\!\perp}}
\newcommand{\expit}{\text{expit}}

\usepackage{xcolor}

\usepackage{authblk}
\usepackage{booktabs}
\usepackage{makecell}

\usepackage[super,sort&compress]{natbib} 
\begin{document}
\title{Covariate Adjustment for the Win Odds: Application to Cardiovascular Outcomes Trials}

\author[1]{Cyrill Scheidegger\footnote{Corresponding author: \href{mailto:cyrill.scheidegger@stat.math.ethz.ch}{cyrill.scheidegger@stat.math.ethz.ch}}}

\author[2]{Simon Wandel}

\author[2]{Tobias M\"utze}

\affil[1]{Seminar for Statistics, ETH Zurich, Switzerland}

\affil[2]{Advanced Quantitative Sciences (AQS), Novartis Pharma AG, Basel, Switzerland}

\maketitle

\begin{abstract}
    Covariate adjustment can enhance precision and power in clinical trials, yet its application to the win odds remains unclear. {The win odds is an extension of the win ratio that counts ties as half a win for the treatment and the control group, respectively.} In their original form, {both the win ratio and the win odds} rely on comparing each individual from the treatment group to each individual from the control group in a pairwise manner, and count the number of wins, losses, and ties from these pairwise comparisons. A priori, it is not clear how covariate adjustment can be implemented for the win odds. 
To address this, we establish a connection between the win odds and the marginal probabilistic index, a measure for which covariate adjustment theory is well-developed.
Using this connection, we show how covariate adjustment for the win odds is possible, leading to potentially more precise estimators and larger power as compared to the unadjusted win odds.
We present the underlying theory for covariate adjustment for the win odds in an accessible way and apply the method on synthetic data based on the CANTOS trial (ClinicalTrials.gov identifier: NCT01327846) characteristics, {on a subset of the HF-ACTION trial data (ClinicalTrials.gov identifier: NCT00047437),} and on simulated data to study the operating characteristics of the method. We observe that there is indeed a potential gain in power when the win odds is adjusted for baseline covariates if the baseline covariates are prognostic for the outcome. This comes at the cost of a slight inflation of the type I error rate for small sample sizes.
\end{abstract}

\section{Introduction}
Composite endpoints are common in cardiovascular clinical trials.\cite{anker2016traditional,lim2008composite}
Examples include the composite of cardiovascular (CV) death and heart failure (HF) hospitalization \citep{packer2020cardiovascular, McMurrayAngiotensin} or the composite of major adverse cardiovascular events (MACE). \cite{ridker2017antiinflammatory, sabatine2017evolocumab, wiviott2019dapagliflozin}
Traditionally, such outcomes are analyzed by fitting a Cox proportional hazards model to the time to the first composite event, ignoring the different clinical importance of death and non-fatal events, e.g., the difference in the clinical importance of CV death and HF hospitalization. In contrast, the win ratio methodology introduced by Pocock et al.\cite{PocockWinRatio} provides a way to reflect the difference in importance of clinical events. The idea of the win ratio methodology is to compare each individual in the treatment group with each individual in the control group, and for each comparison, a ``winner'' is determined according to some pre-specified rule. For the composite of CV death and HF hospitalization, Pocock et al.\cite{PocockWinRatio} proposed the following rule: first, the ``winner'' is determined by having the longer time to CV death. If no CV death occurs during the shared follow-up time, the winner is instead determined based on having the longer time to first HF hospitalization. If the winner still cannot be decided, the comparison is tied. The win ratio is the probability of a ``win'' for the treatment group divided by the probability of a ``loss'' for the treatment group, when the individuals from the treatment and the control group are randomly selected. The win ratio is estimated by $\frac{\text{\# wins}}{\text{\# losses}}$, where $\text{\# wins}$ and $\text{\# losses}$ denote the number of ``wins'' and ``losses'' for the treatment group, when considering all pairwise comparisons between individuals from the treatment and the control group. A win ratio greater than 1 is in favor of the treatment group, and a win ratio less than 1 is in favor of the control group. 

{The application of the win ratio is not limited to composite endpoints of two time-to-event components.\citep{PocockWinRatioInCardiologyTrials, GregsonHierarchical}} The rule for determining a winner can be extended to an arbitrary number of components and, more generally, the methodology can be applied as soon as one has a way to classify the comparisons of outcomes into ``win'', ``loss'', or ``tie''. For example, in the PARACHUTE-HF trial,\citep{bocchi2024sacubitril} the primary endpoint was a hierarchical composite of time to CV death, time to first HF hospitalization, and relative change from baseline to week 12 in NT-proBNP levels. For the latter, a relative change of greater than $25\%$ was required to classify the comparison as a win/loss.  
Due to its broad applicability, the win ratio has gained increasing popularity in the last decade, and a large body of work has been dedicated to developing methods for the statistical analysis (see, for example, Luo et al.,\cite{LuoAnAlternativeApproach} Bebu and Lachin,\cite{BebuLargeSampleInference} Dong et al. \cite{DongAGeneralizedAnalyticSolution}). {For a general overview of the developments in the last years, we refer to the review articles by Pocock et al.\cite{PocockWinRatioInCardiologyTrials} and Gregson et al.\cite{GregsonHierarchical}}

{A common critique of the win ratio is that tied outcomes are not incorporated directly into its definition. When ties occur, the win ratio therefore depends only on the relative frequencies of wins and losses and is unaffected by the proportion of ties. To address this feature, the win odds has been proposed, which adds half the probability of a tie to both the win and the loss probability, yielding the estimator $\frac{\text{\# wins + 0.5 \# ties}}{\text{\# losses + 0.5 \# ties}}$.\citep{DongTheWinRatio, BrunnerWinOdds, SongTheWinOdds} When no ties are present, the win ratio and win odds coincide. In the presence of ties (and if the number of wins is not equal to the number of losses), the win odds pushes the ratio between wins and losses toward 1. Some authors have argued that the win ratio may exaggerate treatment effects when ties are frequent, see Ajufo et al.,\citep{AjufoFallaciesOfUsingTheWR} Butler et al.,\citep{ButlerWRSeductive} and examples therein. Others argue that the win odds is itself difficult to interpret and recommend to rather report the win ratio together with the win difference/net benefit (difference between win and loss probability).\cite{PocockWinRatioInCardiologyTrials, GregsonHierarchical} In addition to this debate, concerns regarding the interpretability and dependence on follow-up time and censoring distribution apply to both methods.\cite{OakesOnTheWinRatio, DongTheWinRatioImpactOfCensoring} This manuscript neither wants to address the debate nor these additional concerns. We focus on the win odds, because it naturally connects to the so-called \textit{marginal probabilistic index} (MPI) $\nu$ studied by Vermeulen et al.,\cite{VermeulenIncreasingThePower} $\nu = \Prob(Y \prec Y^*) + \frac{1}{2} \Prob(Y \asymp Y^*)$, where $Y$ is the outcome of a randomly drawn individual from the control group, $Y^*$ is the outcome of a randomly drawn individual from the treatment group, $Y \prec Y^*$ means that the (randomly drawn) individual from the treatment group ``wins'' against the (randomly drawn) individual from the control group (i.e., outcome $Y^*$ is more favorable than outcome $Y$), and $Y \asymp Y^*$ means that the two individuals are tied.} The quantity $\nu$ is closely related to the underlying effect measure targeted by the Mann-Whitney U test, and it is sometimes also referred to as the \textit{win proportion}, \textit{win probability}, \textit{relative effect}, or \textit{Mann-Whitney parameter}.\citep{BrunnerWinOdds,GasparyanAdjustedWinRatio} The transformation $\frac{\nu}{1-\nu}$ of the MPI $\nu$ then coincides with the win odds.

Adjusting for baseline covariates is common in clinical trials and if the covariates are correlated with the outcome, this can increase statistical power.\cite{ye2023toward, VanLanckerCovariateAdjustment} The guideline on covariate adjustment\cite{FDACovariateGuideline} by the US Food and Drug Administration (FDA) states: ``Therefore, FDA recommends that sponsors adjust for covariates that are anticipated to be most strongly associated with the outcome of interest''. The direct estimators for the win ratio and the win odds are based on counts, so it is not immediately clear how covariate adjustment can be implemented. However, Vermeulen et al.\cite{VermeulenIncreasingThePower} showed how the MPI $\nu$ can be adjusted for covariates using the methodology of probabilistic index models,\cite{ThasProbabilisticIndexModels} leading to estimators with a smaller variance. Then, a covariate adjusted estimator $\hat \nu$ for $\nu$ can be transformed to a covariate adjusted estimator $\frac{\hat \nu}{1-\hat\nu}$ for the win odds with potentially smaller variance than the direct estimator $\frac{\text{\# wins + 0.5 \# ties}}{\text{\# losses + 0.5 \# ties}}$. Despite this simple relationship and the potential advantages, to the best of our knowledge, this way of covariate adjustment for the win odds is not well-known yet in practice and has not been applied to clinical trials. Hence, our goal is to draw attention to this connection and provide guidance for when covariate adjustment can be useful.
More concretely, our main contributions are the following.
\begin{enumerate}
    \item Building on the work from Vermeulen et al.,\cite{VermeulenIncreasingThePower} we present the methodology and relevant theory for covariate adjustment for the MPI and clarify its relationship to the win odds.
    \item We study the operating characteristics for the adjusted win odds for scenarios motivated by cardiovascular outcomes trials in order to provide guidance for clinical trialists.
\end{enumerate}

The remainder of this manuscript is organized as follows. Section \ref{sec_Theory} presents the methodology for covariate adjustment for the MPI, as developed by Vermeulen et al.,\cite{VermeulenIncreasingThePower} and extends it to the win odds. In Section \ref{sec_CantosApplication}, we apply the proposed covariate-adjusted win odds method to synthetic data based on the CANTOS trial\cite{ridker2017antiinflammatory} characteristics {and to a publicly available subset of the HF-ACTION trial data\cite{OConnorHFACTION, MaoWRPackage}}and illustrate the impact of adjusting for different sets of covariates. Section \ref{sec_Simulations} reports results from a simulation study evaluating the operating characteristics of the covariate-adjusted win odds. Finally, Section \ref{sec_Discussion} concludes with a discussion.

\section{Covariate adjustment for the win odds}\label{sec_Theory}
\subsection{Literature review}
Most of the methodology that we use in the following was developed by Vermeulen et al.\cite{VermeulenIncreasingThePower} They consider covariate adjustment for the MPI and focus on numerical outcomes, for which win/loss/tie is just the usual ordering relation. We will use their methodology and apply it for more general definitions of win/loss/tie, allowing for composite endpoints in time-to-event settings. Vermeulen et al.\cite{VermeulenIncreasingThePower} heavily rely on the \textit{probabilistic index model} (PIM) framework, which was introduced by Thas et al.\cite{ThasProbabilisticIndexModels} With a PIM, it is in principle possible to obtain inference for the treatment effect adjusted for covariates, and the connection of PIMs to the win odds has been discussed by Song et al.\cite{SongTheWinOdds} However, the effect measure associated with the treatment in such models is a conditional effect, and correct inference depends on the correct specification of the model. A structurally similar approach for the win ratio was proposed by Mao and Wang.\cite{MaoProportionalWinFractions} They model the ratio of covariate-specific win and loss probabilities as a function of covariates and additionally impose the assumption that the ratio stays constant over time, which has the advantage that the regression coefficients do not depend on the censoring distribution. However, their effect measure is still a conditional effect and not a marginal effect. To our knowledge, the only reference other than Vermeulen et al.\cite{VermeulenIncreasingThePower} that also targets covariate adjustment for the MPI is Gasparyan et al.,\cite{GasparyanAdjustedWinRatio} which give an alternative way to adjust the marginal win odds for a covariate. However, the method of Gasparyan et al. is limited to a univariate numerical covariate, whereas our proposal allows for an arbitrary number of covariates. Wang et al.\cite{WangAdjustedWinRatio} use inverse probability of treatment weighting to adjust the win ratio for baseline imbalances in patients characteristics in the treatment and control group, which is a different objective from ours. Finally, very recently, Cao et al.\cite{CaoCovariateAdjustedWinStatistic} combine inverse probability weighting, overlap weighting, and augmentation to increase the efficiency of win statistics estimators. However, their method is limited to ordinal outcomes with a finite number of categories and, hence, is not directly applicable to the time-to-event setting that we are interested in.

\subsection{Covariate adjustment for the marginal probabilistic index}
We consider a setup similar to Vermeulen et al.:\cite{VermeulenIncreasingThePower} let $(Y_i, A_i, X_i)_{i=1, \ldots, n}$ be independent and identically distributed (i.i.d.). Here, $A_i\in \{0,1\}$ is the randomized treatment and $X_i\in \mathbb R^p$ is a vector of baseline covariates that are independent of the treatment, that is, $A_i\indep X_i$. To ensure this independence, only covariates measured prior to randomization should be included. $Y_i\in \mathbb R^d$ represents the outcome. $Y_i$ can be numerical or ordinal, but also much more general. For example, in the context of a time-to-event analysis with a composite endpoint of CV death and HF hospitalization, we could represent the outcome as $Y_i\in \mathbb R^2 \times \{0,1\}^2$, where the first two components correspond to the time to CV death and HF hospitalization and the last two components indicate whether CV death and HF hospitalization occurred or if they are censored. We further assume that we have a ``win/loss/tie rule'' such that for each pair $(Y_i, Y_j)$, exactly one of the following holds: i) $Y_i \prec Y_j$ (i.e., {individual $j$ wins against individual $i$}), ii) $Y_j \prec Y_i$ (i.e., {individual $i$ wins against individual $j$}), iii) $Y_i \asymp Y_j$ (i.e., {individuals $i$ and $j$ are tied}). In Vermeulen et al.,\cite{VermeulenIncreasingThePower} the motivation was covariate adjustment for the Mann-Whitney U test, and therefore $Y_i$ was assumed to be univariate and numerical, and the relation $\prec$ corresponded to $<$ and $\asymp$ corresponded to $=$. However, our setting is more general: for example, the win/loss/tie relation for the time-to-event setting outlined in the introduction is not transitive and, hence, cannot be transformed to a numerical outcome with standard ordering relation. Hence, we impose no restrictions on the win/loss/tie rule other than that each pair of observations can be compared.

The parameter of interest from Vermeulen et al.\cite{VermeulenIncreasingThePower} is the marginal probabilistic index (MPI)
$$\nu = \Prob(Y_i\preceq Y_j|A_i = 0, A_j = 1) \coloneqq \Prob(Y_i \prec Y_j|A_i = 0, A_j = 1) + \frac{1}{2}\Prob(Y_i \asymp Y_j|A_i = 0, A_j = 1),$$
that is, the probability that a randomly chosen individual from the treatment group wins against a randomly chosen individual from the control group plus half the probability of a tie. The win odds is defined as the odds of the MPI, that is,
$$\theta = \frac{\nu}{1-\nu} = \frac{\Prob(Y_i \prec Y_j|A_i = 0, A_j = 1) + \frac{1}{2}\Prob(Y_i \asymp Y_j|A_i = 0, A_j = 1)}{\Prob(Y_i \succ Y_j|A_i = 0, A_j = 1) + \frac{1}{2}\Prob(Y_i \asymp Y_j|A_i = 0, A_j = 1)}.$$
Let $N_1 = \sum_{i=1}^n A_i$ be the number of units in the treatment group and $N_0 = \sum_{i=1}^n (1-A_i)$ be the number of units in the control group. Moreover, define $I(Y_i\preceq Y_j) = I(Y_i\prec Y_j) + \frac{1}{2}I(Y_i \asymp Y_j)$, where $I(Y_i\prec Y_j)$ and $I(Y_i \asymp Y_j)$ are the usual indicator functions. Then, a straightforward estimator for $\nu$ is the number of wins plus half the number of ties divided by the total number of comparisons $N_0 N_1$, i.e., 
$$\hat\nu_\text{direct} = \frac{1}{N_0 N_1}\sum_{i=1}^n \sum_{j\neq i}(1-A_i)A_jI(Y_i \preceq Y_j),$$
and the direct estimator $\frac{\text{\# wins + 0.5 \# ties}}{\text{\# losses + 0.5 \# ties}}$ for the win odds is equal to $\hat \theta_\text{direct} = \frac{\hat\nu_\text{direct}}{1-\hat\nu_\text{direct}}$.

Vermeulen et al.\cite{VermeulenIncreasingThePower} show how one can adjust for the covariates $X_i$ in order to construct a more efficient estimator for $\nu$. Such an estimator will automatically give rise to a more efficient estimator for the win odds $\theta$. The estimator introduced by Vermeulen et al.\cite{VermeulenIncreasingThePower} can be motivated from two perspectives.

The first perspective is based on the fact that
$$\nu = \Prob(Y_i\preceq Y_j|A_i = 0, A_j = 1) = \E_{(X_i, X_j)}\left[\Prob(Y_i\preceq Y_j|A_i = 0, A_j = 1, X_i, X_j)\,\big|\,A_i = 0, A_j = 1\right],$$
where $\E_{(X_i, X_j)}[\cdot]$ is the expectation taken over the distribution of $(X_i, X_j)$.
Hence, in the first step, one estimates the so-called conditional probabilistic index (CPI) $\Prob(Y_i\preceq Y_j|A_i, A_j, X_i, X_j)$ and marginalizes over the covariates in the second step. Concretely, one postulates a model
\begin{equation}
    \Prob(Y_i\preceq Y_j|A_i, A_j, X_i, X_j) = m(A_i, A_j, X_i, X_j),\label{eq_CPI}
\end{equation}
estimates $m(\cdot)$ by $\hat m(\cdot)$, and uses standardization \citep{VansteelandtGComputation, VanLanckerCovariateAdjustment} to obtain
\begin{equation}
    \hat\nu_\text{stand.} = \frac{1}{n(n-1)}\sum_{i=1}^n\sum_{j\neq i}\hat m(0, 1, X_i, X_j).\label{eq_Standardization}
\end{equation}
The estimator $\hat m$ is obtained by postulating a probabilistic index model (PIM),\citep{ThasProbabilisticIndexModels} that is, $m(\cdot)$ is modeled by
\begin{equation}
    m(A_i, A_j, X_i, X_j) = g^{-1}\left(\tau_A (A_j - A_i) + \tau_X^T(X_j - X_i)\right),\label{eq_PIM}
\end{equation}
with a fixed link function $g: (0,1)\to \mathbb R$, $\tau_A\in \mathbb R$, and $\tau_X\in \mathbb R^p$, where $\tau_X^T$ denotes transposition of the vector $\tau_X$. Hence, the function $m(\cdot)$ only depends on the treatments $A_i$, $A_j$ and the covariates $X_i$, $X_j$ through the differences $A_j - A_i$ and $X_j - X_i$, and the dependence is modeled using the pre-specified link function $g(\cdot)$. Then, the estimator $\hat m$ is obtained by plugging in estimators $\hat\tau_A$ and $\hat \tau_X$ for $\tau_A$ and $\tau_X$.
We will exclusively focus on the logit link function $g(t) = \log\left(\frac{t}{1-t}\right)$, but other link functions such as, for example, the probit or identity link are possible. It is also possible to include interactions by adding a term $\tau_{AX}^T(A_j X_j - A_i X_i)$ to \eqref{eq_PIM} and similarly, arbitrary transformations of $X_i$ can be included, but we do not discuss this further here. To summarize, with the logit link function, the CPI \eqref{eq_CPI} is modeled as
$$\Prob(Y_i\preceq Y_j|A_i, A_j, X_i, X_j) = \expit\left(\tau_A (A_j - A_i) + \tau_X^T(X_j - X_i)\right)$$
with $\expit(t) = \frac{\exp(t)}{1 + \exp(t)}$.
Estimators $\hat \tau_A$ and $\hat \tau_X$ of the parameters $\tau_A$ and $\tau_X$ can be obtained by solving the estimating equations\cite{ThasProbabilisticIndexModels, JaspersCovariateAdjustedGPC}
\begin{equation}
    \sum_{i=1}^n\sum_{j\neq i} \binom{A_j - A_i}{X_j - X_i}\left(I(Y_i \preceq Y_j) - \expit\left(\hat \tau_A (A_j - A_i) + \hat \tau_X^T(X_j - X_i)\right)\right) = 0\in \mathbb R^{1+p},\label{eq_EstEqPIM}
\end{equation}
where $\binom{A_j - A_i}{X_j - X_i}$ is a column vector in $\mathbb R^{1+p}$. 
Equation \eqref{eq_EstEqPIM} is equivalent to the logistic regression estimating equations in the $n\cdot(n-1)$ ``pseudo-observations'' $\left\{I(Y_i \preceq Y_j)\right\}_{i = 1,\ldots, n,\, j\neq i}$. \cite{ThasProbabilisticIndexModels} However, due to the nontrivial dependence structure of the pseudo-observations, there might exist more efficient methods to estimate $\tau_A$ and $\tau_X$.\cite{ThasProbabilisticIndexModels, VermeulenSemiparametricEstimation}
Nevertheless, a practical advantage of using the estimating equations \eqref{eq_EstEqPIM} is that $\hat \tau_A$ and $\hat \tau_X$ can be obtained using standard software for logistic regression (see also Section \ref{sec_PracticalImplementation} below). 
Putting everything together, we can rewrite $\hat \nu_\text{stand.}$ in \eqref{eq_Standardization} as
\begin{equation}
    \hat\nu_\text{stand.} = \frac{1}{n(n-1)}\sum_{i=1}^n\sum_{j\neq i}\expit\left(\hat \tau_A + \hat\tau_X^T(X_j - X_i)\right).\label{eq_NuStand}
\end{equation}
This first way to motivate the covariate adjustment for the MPI via standardization is very intuitive. However, from \eqref{eq_NuStand}, it is not clear how inference for $\hat \nu_\text{stand.}$ can be carried out. Moreover, postulating the PIM \eqref{eq_PIM} seems like a strong modeling assumption which might not be satisfied in practice.

These questions can be resolved by looking at the second way of motivating the covariate adjustment. Vermeulen et al.\cite{VermeulenIncreasingThePower} show that for an arbitrary function $H(X_i, X_j)$, the estimator
\begin{equation}
    \hat\nu(H) = \hat\nu_\text{direct} + \sum_{i=1}^n \sum_{j\neq i}\left(\frac{1}{n(n-1)}-\frac{(1-A_i)A_j}{N_0N_1}\right)H(X_i, X_j)\label{eq_AdjArbitrary}
\end{equation}
is an asymptotically normal estimator of $\nu$. The idea is to add an augmentation term to $\hat\nu_\text{direct}$ that has (asymptotically) mean zero but influences the variance of the estimator. Hence, one can look for the function $H(X_i, X_j)$ that leads to the smallest asymptotic variance. Vermeulen et al.\cite{VermeulenIncreasingThePower} show that the choice
\begin{equation}
    H^*(X_i, X_j) = \Prob(Y_i\preceq Y_j|A_i = 0, A_j = 1, X_i, X_j)\label{eq_Hstar}
\end{equation}
leads to the smallest variance and achieves the semiparametric efficiency bound. Again, one can postulate a PIM to estimate this conditional probability. This leads to the estimator
\begin{equation}
    \hat\nu_\text{aug.} = \hat\nu(\hat H^*) = \hat\nu_\text{direct} + \sum_{i=1}^n \sum_{j\neq i}\left(\frac{1}{n(n-1)}-\frac{(1-A_i)A_j}{N_0N_1}\right)\expit\left(\hat \tau_A + \hat\tau_X^T(X_j - X_i)\right),\label{eq_NuAug}
\end{equation}
where $\hat \tau_A$ and $\hat \tau_X$ are estimated by \eqref{eq_EstEqPIM}. Vermeulen et al.\cite{VermeulenIncreasingThePower} show that the standardization estimator from \eqref{eq_NuStand} is equal to this augmentation estimator from \eqref{eq_NuAug}, that is, $\hat \nu_\text{stand.} = \hat \nu_\text{aug.}$. This property exclusively holds for the logit link with the estimating equations \eqref{eq_EstEqPIM}. However, since $\hat\nu_\text{aug.}$ resembles a U-statistic, it is relatively straightforward to estimate its variance using a sandwich estimator.\cite{VermeulenIncreasingThePower} Moreover, since \eqref{eq_AdjArbitrary} produces a valid estimator for arbitrary functions $H(X_i, X_j)$, it is also evident that misspecification of the PIM is not a severe problem. It only affects the efficiency of the estimator, but does not invalidate the inference.

In summary, when modeling the CPI using a PIM with logit link and using the estimating equations \eqref{eq_EstEqPIM}, the estimators \eqref{eq_NuStand} and \eqref{eq_NuAug} are identical. If the PIM is correctly specified, the estimator is semiparametrically efficient. If not, it is still consistent and asymptotically normal. For more details on the theory, we refer to Vermeulen et al.\cite{VermeulenIncreasingThePower}

\subsection{Inference for the adjusted win odds}\label{sec_InferenceAdjWO}
Hereinafter, we write $\hat\nu$ for the estimator $\hat \nu_\text{stand.}= \hat\nu_\text{aug.}$. The variance estimator is given in equation (A8) in Vermeulen et al.\cite{VermeulenIncreasingThePower} Denote this variance estimator by $\hat\sigma_\nu^2$. Then, the Wald statistic $(\hat\nu - \nu)/\hat\sigma_\nu$ converges in distribution to a standard normal distribution, that is, $(\hat\nu - \nu)/\hat\sigma_\nu\to\mathcal N(0, 1)$. A p-value for the null hypothesis of no treatment effect can be obtained from the asymptotic normality of the test statistic $(\hat\nu - 0.5)/\hat \sigma_\nu$ under the null hypothesis. Similarly, an asymptotic $(1-\alpha)$-confidence interval for $\nu$ can be obtained by $\hat\nu \pm \Phi^{-1}(1-\alpha/2) \hat\sigma_\nu$, where $\Phi(\cdot)$ is the cumulative distribution function of a standard normal random variable.

To transform the inference to the win odds scale, one can use the estimator $\hat \theta = \frac{\hat\nu}{1-\hat \nu}$ for the win odds. The confidence interval can also be forward transformed, that is, if $[C_l,\, C_r]$ is an asymptotic $(1-\alpha)$ confidence interval for $\nu$, then $[C_l/(1-C_l),\, C_r/(1-C_r)]$ is an asymptotic $(1-\alpha)$-confidence interval for the win odds $\theta$ since $\nu\mapsto \nu/(1-\nu)$ is monotonically increasing. Since the null hypothesis $H_0: \nu = 0.5$ is equivalent to the null hypothesis $H_0: \theta = 1$, the same p-value can be used for the win odds $\theta$ as for the MPI $\nu$. Alternatively, one can use the delta method for $\log\hat \theta$ and obtain an estimator $\hat\sigma^2_{\log\theta} = \hat\sigma_\nu^2/(\hat\nu(1-\hat\nu))^2$ of the variance of $\log\hat \theta$ and construct tests and confidence intervals based on the asymptotic normality of $\log\hat\theta$.

Since the tests are based on a normal approximation, there can be a slight type I error inflation for small samples (see also the simulations in Section \ref{sec_Simulations}). For this reason, Vermeulen et al.\cite{VermeulenIncreasingThePower} also introduce a permutation test procedure based on the estimator $\hat\nu$ (see Section 6 there). Since our main interest in this manuscript lies in the application to cardiovascular outcomes trials, where the sample size is usually large, we do not consider this approach further.

\subsection{Practical implementation}\label{sec_PracticalImplementation}
R code to calculate the marginal probabilistic index (based on equations \eqref{eq_NuStand} and \eqref{eq_NuAug}) with variance, p-values, and confidence intervals is provided in the Web Appendix E of Vermeulen et al.\cite{VermeulenIncreasingThePower} Their code assumes a numerical outcome $Y_i$ and uses the R package \texttt{pim}\cite{PIMPackage} available on CRAN to obtain the coefficients of the PIM. Hence, it is not exactly applicable to our setting which allows for arbitrary definitions of wins/losses/ties. For the applications that follow in Section \ref{sec_CantosApplication} and \ref{sec_Simulations}, we calculate the win/loss/tie outcome for every pair $i \neq j$, $i,j = 1,\ldots, n$ and obtain the coefficients of the PIM by solving \eqref{eq_EstEqPIM} using the function \texttt{glm()} from the \texttt{stats} R package\citep{RCoreTeamR} or the function \texttt{bigglm()} from the R package \texttt{biglm}\citep{LumleyBIglm} for large sample sizes. Given the coefficients of the PIM, one can then follow the R code from Vermeulen et al.\cite{VermeulenIncreasingThePower} to obtain the MPI estimator and its standard error, the p-value and the confidence interval. These can be transformed back to the win odds scale as outlined in Section \ref{sec_InferenceAdjWO}. Our implementation, which heavily uses large parts of the code provided by Vermeulen et al.,\cite{VermeulenIncreasingThePower} can be requested from the authors.

\section{{Application to cardiovascular outcomes trials}}
\subsection{Application to synthetic CANTOS trial data}\label{sec_CantosApplication}
We apply the proposed method for the covariate-adjusted win odds to synthetic CANTOS trial data.\citep{ridker2017antiinflammatory} 
In the CANTOS trial, the efficacy and safety of canakinumab compared to placebo were assessed in patients with a prior myocardial infarction and elevated high-sensitivity C-reactive protein (hsCRP). The primary endpoint was the time to the first occurrence of a composite of nonfatal myocardial infarction (MI), nonfatal stroke, or cardiovascular death (three-point MACE). In the original trial, 10\,061 patients were randomly assigned to receive canakinumab (at doses of 50 mg, 150 mg, or 300 mg every three months) or placebo, in addition to standard therapy. The primary analysis was performed using a Cox proportional hazards model for the time to the first event of the composite outcome, stratified according to the time since the index myocardial infarction and according to trial part.

We use synthetic data based on the CANTOS trial characteristics for evaluating our statistical methods. When creating the synthetic dataset, a model is built to capture the distributions and relationships in the original dataset. Then, this model is used to generate a realization of synthetic data. While these synthetic data do not directly match any original records, they retain the key statistical properties of the CANTOS data. Table \ref{tab_CantosSynDesc} summarizes the number of events for different clinical outcomes in the synthetic dataset.

\begin{table}
\centering
\scriptsize
\begin{tabular}[t]{lrrrr}
\toprule
Clinical outcome & \makecell{Placebo\\($n=3344$)} & \makecell{Treatment 50 mg\\($n=2170$)} & \makecell{Treatment 150 mg\\($n=2284$)} & \makecell{Treatment 300 mg\\($n=2263$)}\\
\midrule
MACE & 535 & 313 & 320 & 322\\
CV Death or death of unknown cause & 235 & 137 & 144 & 151\\
Non-fatal MI & 291 & 168 & 158 & 171\\
Non-fatal Stroke & 91 & 58 & 63 & 51\\
\bottomrule
\end{tabular}
\caption{Number of events for clinical outcomes in the synthetic CANTOS trial data.}
\label{tab_CantosSynDesc}
\end{table}

We analyze the synthetic CANTOS trial data using the win odds. To compare the outcomes of individuals $i$ and $j$, we use a win/loss/tie rule that is similar to the rule from Pocock et al.\cite{PocockWinRatio}:
\begin{enumerate}
    \item If there was at least one CV death during the shared follow-up time of $i$ and $j$, the individual for which the CV death occurred later (or never) wins.
    \item If there was no CV death and at least one non-fatal MI or non-fatal stroke during the shared follow-up time of $i$ and $j$, the individual for which the first non-fatal MI or non-fatal stroke occurred later (or never) wins.
    \item If the winner cannot be determined on the basis of 1. or 2., the comparison is tied.
\end{enumerate}
In the adjusted win odds, we consider adjusting for the following covariates:
\begin{description}
    \item[Time since the index MI:] The analysis of the CANTOS trial\cite{ridker2017antiinflammatory} adjusted for the time since the index MI; hence, it is natural to also consider this variable as an adjustment variable here. Time since the index MI is a categorical variable with two levels ($\geq$6 months, $<$6 months). 
    \item[log hsCRP at baseline:] One of the inclusion criteria of the study was an hsCRP level of at least 2 mg/L.
    \item[Dichotomized hsCRP at baseline:] Instead of taking the logarithm of the hsCRP baseline value, we take an indicator if the value is higher than or lower than the median hsCRP baseline value. Values equal to the median hsCRP baseline value are part of to the `lower than' group. 
\end{description}
In the synthetic CANTOS trial data, the covariate values are available for all subjects. We use the adjustment sets \{time since the index MI\}, \{log hsCRP at baseline\}, \{dichotomized hsCRP at baseline\}, \{time since the index MI\, dichotomized hsCRP at baseline\}, and \{time since the index MI, dichotomized hsCRP at baseline\}. We use the implementation described in Section \ref{sec_PracticalImplementation} for the adjusted win odds. Inference for the unadjusted win odds is based on the derivations in Bebu and Lachin\cite{BebuLargeSampleInference} and the delta method. To be precise, Bebu and Lachin\cite{BebuLargeSampleInference} provide inference for the \textit{proportion in favor parameter} $\Delta$, i.e., the probability of a win minus the probability of a loss. The delta method can be applied because the win odds $\theta = \frac{1 + \Delta}{1-\Delta}$. Finally, we also use the adjustment from Gasparyan et al.,\cite{GasparyanAdjustedWinRatio} which is only applicable for a single numerical covariate, and we use the formulas given in Section 3.1 there. The results are shown in Table \ref{tab_SynCantosResults300mg}. 

\begin{table}[!h]
\centering
\scriptsize
\begin{tabular}{llccc}
\toprule
Adjustment set & Statistic & unadjusted & adjusted & univ. adjusted\\
\midrule
\{Time since the index MI\} & WO & 1.041263 & 1.041073 & 1.041100\\
 & 95\% CI & (1.0056, 1.0782) & (1.0055, 1.0780) & (1.0055, 1.0780)\\
 & p-value & 0.022826 & 0.023426 & 0.023314\\
\addlinespace
\{log hsCRP at baseline\} & WO & 1.041263 & 1.043316 & 1.043393\\
 & 95\% CI & (1.0056, 1.0782) & (1.0077, 1.0802) & (1.0078, 1.0803)\\
 & p-value & 0.022826 & 0.016611 & 0.016405\\
\addlinespace
\{Dichotomized hsCRP at baseline\} & WO & 1.041263 & 1.041528 & 1.041527\\
 & 95\% CI & (1.0056, 1.0782) & (1.0060, 1.0784) & (1.0060, 1.0784)\\
 & p-value & 0.022826 & 0.021732 & 0.021713\\
\addlinespace
\{Time since the index MI, log hsCRP at baseline\} & WO & 1.041263 & 1.043157 & -\\
 & 95\% CI & (1.0056, 1.0782) & (1.0076, 1.0800) & -\\
 & p-value & 0.022826 & 0.017004 & -\\
\addlinespace
\{Time since the index MI, Dichotomized hsCRP at baseline\} & WO & 1.041263 & 1.041362 & -\\
 & 95\% CI & (1.0056, 1.0782) & (1.0058, 1.0782) & -\\
 & p-value & 0.022826 & 0.022249 & -\\
\bottomrule
\end{tabular}
\caption{Results for unadjusted, adjusted (using the methodology of Vermeulen et al.\cite{VermeulenIncreasingThePower}), and univariate adjusted \citep{GasparyanAdjustedWinRatio} win odds on the 300 mg treatment group for the synthetic CANTOS trial data. The p-values are two-sided.}
\label{tab_SynCantosResults300mg}
\end{table}

We see that the unadjusted and adjusted win odds and confidence intervals are similar across all the different adjustment sets. The p-values for the adjusted win odds are smaller than for the unadjusted win odds for all adjustment sets except for when only adjusting for time since the index MI. The magnitude of the difference is the largest when adjusting for log hsCRP at baseline (either alone or in addition to time since the index MI), and the smallest when adjusting only for dichotomized hsCRP at baseline in combination with time since the index MI. 
For the two adjustment sets where univariate adjustment according to Gasparyan et al.\cite{GasparyanAdjustedWinRatio} is possible, the results are almost identical to the adjustment using the methodology based on Vermeulen et al.\cite{VermeulenIncreasingThePower}
It is worth noting that the scenarios with a lower p-value also have a larger effect size. The length of the confidence intervals confirms that in the settings studied here, there is no noticeable efficiency gain by adjusting for covariates and that differences in the p-values are primarily driven by differences in the point estimates. 
Results for the other treatment groups are shown in Tables \ref{tab_SynCantosResults150mg} and \ref{tab_SynCantosResults50mg}. The results are qualitatively similar.

{
\subsection{Application to a subset of the HF-ACTION trial data}
To complement the synthetic CANTOS trial data example, we additionally apply the covariate-adjusted win odds to a publicly available subset of the HF-ACTION trial data,\cite{OConnorHFACTION} comprising 451 non-ischemic heart failure patients, available in the \texttt{WR} R package.\cite{MaoWRPackage} The trial compared exercise training to usual care, with a composite endpoint of all-cause death and hospitalization. Following the analysis in the vignette on proportional win-fractions regression\cite{MaoProportionalWinFractions} in the \texttt{WR} package,\cite{MaoWRVignette} we can adjust for the following covariates: age, male vs. female, black/other vs. white, BMI, left-ventricular ejection fraction, hypertension, COPD, diabetes, ACE inhibitor use, beta-blocker use, and current smoking status. We consider adjusting for each available baseline covariate individually as well as adjusting for all covariates simultaneously. The results are shown in Table \ref{tab_HFACTIONResults}. For some covariates, such as male vs. female and hypertension, covariate adjustment yields modestly smaller p-values, whereas for others, including age, left-ventricular ejection fraction, COPD, and adjustment for all covariates jointly, the adjusted analysis does not improve significance and in some cases leads to slightly larger p-values. Moreover, the changes in p-values are accompanied by small shifts in the point estimate, while the confidence interval widths are only modestly affected. Overall, this real-data example illustrates that covariate adjustment for the win odds is practically feasible in cardiovascular outcome data, but also that meaningful efficiency gains should only be expected when the adjustment variables are sufficiently prognostic for the outcome and are incorporated in a suitably informative way.

\begin{table}[!h]
\centering
\scriptsize
\begin{tabular}{llccc}
\toprule
Adjustment set & Statistic & unadjusted & adjusted & univ. adjusted\\
\midrule
\{Age\} & WO & 1.195580 & 1.188828 & 1.189019\\
 & 95\% CI & (0.9784, 1.4609) & (0.9738, 1.4564) & (0.9744, 1.4560)\\
 & p-value & 0.080681 & 0.089385 & 0.088373\\
\addlinespace
\{Male vs. Female\} & WO & 1.195580 & 1.213929 & 1.213648\\
 & 95\% CI & (0.9784, 1.4609) & (0.9943, 1.4879) & (0.9946, 1.4868)\\
 & p-value & 0.080681 & 0.056904 & 0.056647\\
\addlinespace
\{Black.vs.White, Other.vs.White\} & WO & 1.195580 & 1.196175 & -\\
 & 95\% CI & (0.9784, 1.4609) & (0.9807, 1.4642) & -\\
 & p-value & 0.080681 & 0.077176 & -\\
\addlinespace
\{BMI\} & WO & 1.195580 & 1.191534 & 1.191639\\
 & 95\% CI & (0.9784, 1.4609) & (0.9765, 1.4591) & (0.9770, 1.4585)\\
 & p-value & 0.080681 & 0.084485 & 0.083637\\
\addlinespace
\{LVEF\} & WO & 1.195580 & 1.187675 & 1.187568\\
 & 95\% CI & (0.9784, 1.4609) & (0.9740, 1.4533) & (0.9743, 1.4525)\\
 & p-value & 0.080681 & 0.089377 & 0.088828\\
\addlinespace
\{Hypertension\} & WO & 1.195580 & 1.200550 & 1.200229\\
 & 95\% CI & (0.9784, 1.4609) & (0.9835, 1.4709) & (0.9837, 1.4698)\\
 & p-value & 0.080681 & 0.072462 & 0.072241\\
\addlinespace
\{COPD\} & WO & 1.195580 & 1.185958 & 1.186183\\
 & 95\% CI & (0.9784, 1.4609) & (0.9720, 1.4520) & (0.9726, 1.4516)\\
 & p-value & 0.080681 & 0.093054 & 0.091958\\
\addlinespace
\{Diabetes\} & WO & 1.195580 & 1.191731 & 1.191297\\
 & 95\% CI & (0.9784, 1.4609) & (0.9761, 1.4602) & (0.9762, 1.4590)\\
 & p-value & 0.080681 & 0.085133 & 0.085077\\
\addlinespace
\{ACE Inhibitor\} & WO & 1.195580 & 1.195872 & 1.195924\\
 & 95\% CI & (0.9784, 1.4609) & (0.9794, 1.4656) & (0.9798, 1.4650)\\
 & p-value & 0.080681 & 0.079285 & 0.078544\\
\addlinespace
\{Beta Blocker\} & WO & 1.195580 & 1.193200 & 1.193381\\
 & 95\% CI & (0.9784, 1.4609) & (0.9774, 1.4619) & (0.9780, 1.4614)\\
 & p-value & 0.080681 & 0.082744 & 0.081795\\
\addlinespace
\{Smoker\} & WO & 1.195580 & 1.196006 & 1.195487\\
 & 95\% CI & (0.9784, 1.4609) & (0.9794, 1.4658) & (0.9794, 1.4645)\\
 & p-value & 0.080681 & 0.079159 & 0.079210\\
\addlinespace
\{All of the above\} & WO & 1.195580 & 1.175784 & -\\
 & 95\% CI & (0.9784, 1.4609) & (0.9683, 1.4322) & -\\
 & p-value & 0.080681 & 0.102276 & -\\
\bottomrule
\end{tabular}
\caption{{Results for unadjusted, adjusted (using the methodology of Vermeulen et al.\cite{VermeulenIncreasingThePower}), and univariate adjusted \citep{GasparyanAdjustedWinRatio} win odds on the HF-ACTION trial data subset. The p-values are two-sided.}}
\label{tab_HFACTIONResults}
\end{table}
}

\section{Simulation study: assessing the operating characteristics}\label{sec_Simulations}

Section \ref{sec_CantosApplication} showed that for a particular data set, there might not be efficiency gains. Therefore, in this section, we systematically study the operating characteristics of the covariate-adjusted win odds through simulation studies. We simulate synthetic data that allows us to adjust for a varying degree of covariate information and again assess the power and type I error rate properties.

\begin{figure}[ht]
\centering
\includegraphics[width=0.9\textwidth]{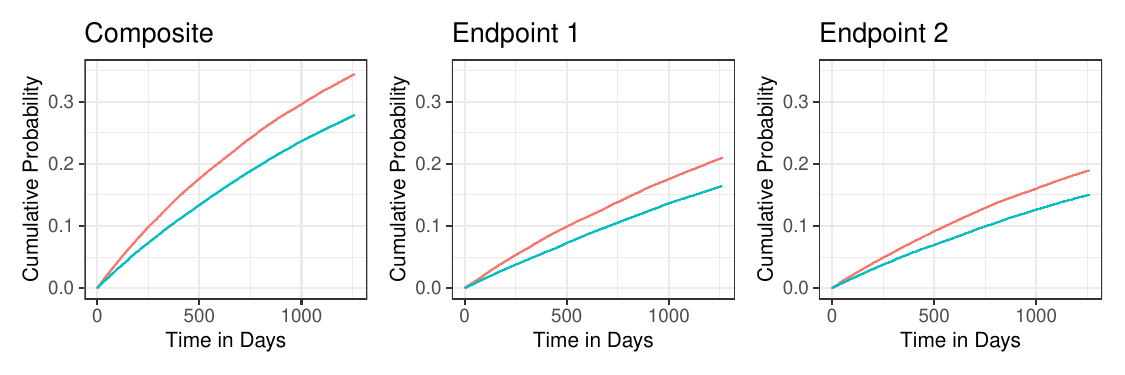}
\caption{Cumulative incidence curves for the composite and for $T_1$ and $T_2$ for one realization of the data simulated according to \eqref{eq_SimSetup}, based on 100\,000 observations of the form \eqref{eq_ObservedData}.}
\label{fig_KMSimulated}
\end{figure}

\begin{figure}[ht]
\centering
\includegraphics[width=0.99\textwidth]{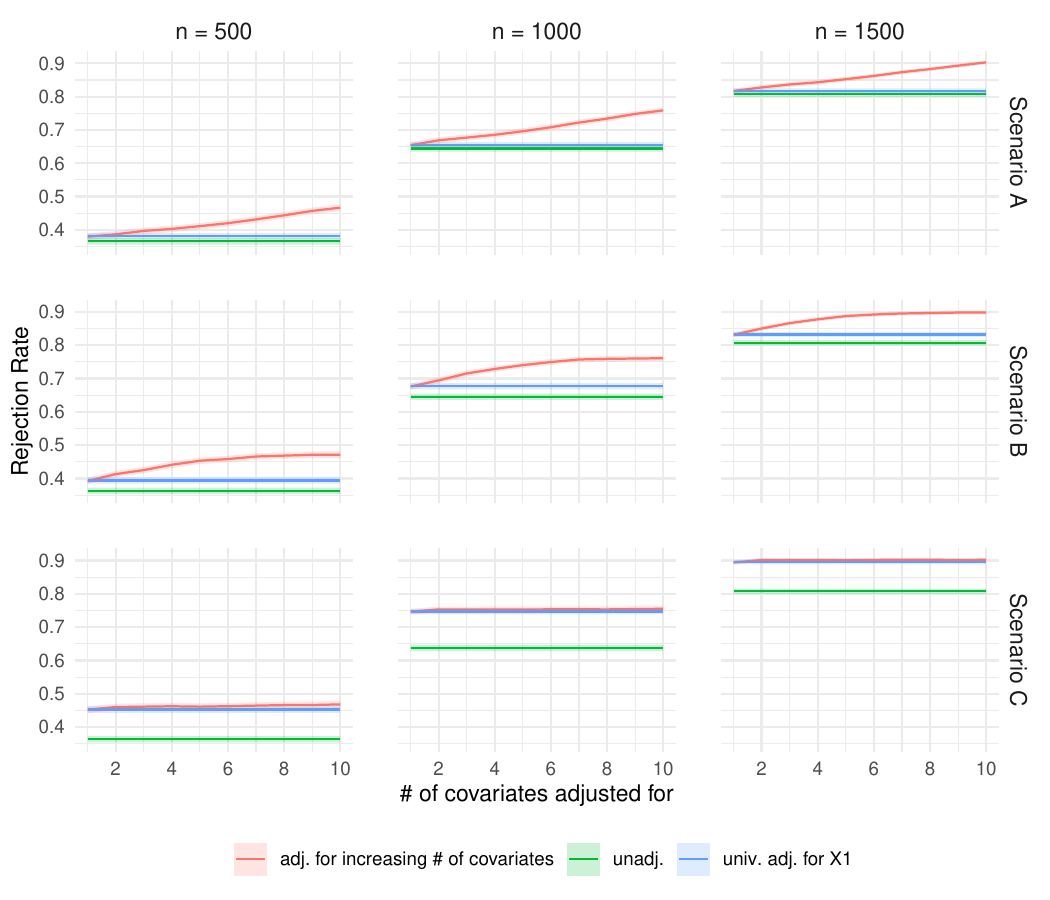}
\caption{Plots of the probability of rejecting the null hypothesis $H_0:\theta\leq 1$ at significance level $\alpha = 0.025$ with datasets of size $n \in \{500, 1000, 1500\}$ generated according to the simulation setup \eqref{eq_SimSetup} and scenarios A, B, and C for the influence of the covariates. The red line corresponds to adjusting for an increasing number of covariates, i.e., adjusting for $\{X_1\}$, $\{X_1, X_2\},\ldots, \{X_1, \ldots, X_{10}\}$, the green line to no adjustment and the blue line to adjusting for $X_1$ using the method from Gasparyan et al.\cite{GasparyanAdjustedWinRatio} Shaded regions indicate 95\% confidence intervals for the estimated rejection probabilities based on 10\,000 replications.}
\label{fig_SimulationHA}
\end{figure}

The simulation setup is designed in such a way that we are able to adjust for a varying degree of covariate information, but the marginal distributions of the other variables in the model stay constant. We generate data according to a latent failure time model as follows:
\begin{align}
    X_1, \ldots, X_{10}&\overset{i.i.d.}{\sim}\mathcal N(0,1),\nonumber\\
    A&\overset{\phantom{i.i.d.}}{\sim} \text{Bernoulli}(0.5),\nonumber\\
    T_1^0, T_2^0&\overset{i.i.d.}{\sim} \text{Exponential}(1), \label{eq_SimSetup}\\
    T_1 &\overset{\phantom{i.i.d.}}{\gets} 7500 \cdot \exp\left(0.3A +\gamma_1X_1+\ldots + \gamma_{10} X_{10}\right)\cdot T_1^0,\nonumber\\
    T_2 &\overset{\phantom{i.i.d.}}{\gets} 7500 \cdot \exp\left(0.3A +\gamma_1X_1+\ldots + \gamma_{10} X_{10}\right)\cdot T_2^0.\nonumber
\end{align}
That is, we assume that conditionally on the treatment assignment and the covariates, both event times $T_1$ and $T_2$ follow an exponential distribution that is consistent with a proportional hazards model linear in the treatment $A$ and the covariates $X_1,\ldots, X_{10}$. We simulate $n\in \{500, 1000, 1500\}$ i.i.d. observations from this model, but we consider censored observations: we choose a censoring time $T_\text{cens.}$ such that 35\% of the individuals experienced at least one event, i.e., $T_\text{cens.} = \hat F_n^{-1}(0.35)$, where $\hat F_n(\cdot)$ is the empirical distribution function of $\min(T_1, T_2)$. Then, only events up to time $T_\text{cens.}$ are observed, and $T_2$ is only observed if it is smaller than $T_1$. To summarize, the outcomes that are observed have the form
\begin{equation}\label{eq_ObservedData}
    Y = \left(\min(T_1, T_\text{cens.}),\, \min(T_2, T_1, T_\text{cens.}),\, I(T_1 < T_\text{cens.}), I(T_2 < \min(T_1, T_\text{cens.})\right).
\end{equation}
We take the same win/loss/tie rule as defined in the previous section with $T_1$ taking the role of time to death and $T_2$ the role of time to first non-fatal MI or non-fatal stroke. In the following, we will explore different choices for the coefficients $\gamma_1, \ldots, \gamma_{10}$ of the covariates. 
\begin{description}
    \item[A:] $\gamma_j = 1/\sqrt{10}, \, j = 1,\ldots, 10$, that is, $X_1, \ldots, X_{10}$ have equal influence on the outcome.
    \item[B:] $\delta_j = (1-(j-1)/10)$, $\gamma_j = \delta_j/\sqrt{\delta_1^2 + \ldots + \delta_{10}^2},\, j = 1,\ldots, 10$, that is, $X_1, \ldots, X_{10}$ have a linearly decreasing influence on the outcome.
    \item[C:] $\delta_j = 1/j^2,\, j = 1,\ldots, 5$ and $\delta_6, \ldots, \delta_{10} = 0$, $\gamma_j = \delta_j/\sqrt{\delta_1^2 + \ldots + \delta_{10}^2},\, j = 1,\ldots, 10$, that is, only $X_1, \ldots, X_5$ have any influence on the outcome and the influence of $X_j$ on the outcome rapidly decreases with $j$.
\end{description}
Note that for all scenarios, $\gamma_1^2 + \ldots + \gamma_{10}^2=1$, so that $\gamma_1 X_1 + \ldots + \gamma_{10}X_{10}\sim \mathcal N(0,1)$, and hence, the marginal distribution of the outcomes is the same across the three scenarios. Cumulative incidence curves (based on the observed quantities \eqref{eq_ObservedData}) for the composite event and for $T_1$ and $T_2$ are shown in Figure \ref{fig_KMSimulated}.
The parameters in \eqref{eq_SimSetup} are chosen in such a way that the cumulative incidence curves of the composite event and of $T_1$ and $T_2$ are comparable to what can typically be observed in cardiovascular outcomes trials.
We calculate empirical rejection rates of the one-sided null hypothesis $H_0:\theta\leq 1$ at significance level $\alpha = 0.025$ based on $10\,000$ independent realizations of the dataset of size $n\in \{500, 1000, 1500\}$. We vary the set of covariates for which we adjust, i.e., we adjust for $\{X_1\}$, $\{X_1, X_2\}$, $\{X_1, X_2, X_3\}, \ldots, \{X_1, \ldots, X_{10}\}$. We compare the rejection rates to the test based on the unadjusted win ratio and the univariate adjustment using the approach by Gasparyan et al.,\cite{GasparyanAdjustedWinRatio} where we always adjust for $X_1$. 
The R code for the simulation study can be requested from the authors. The results are presented in Figure \ref{fig_SimulationHA}.
Looking at the first two rows (scenarios A and B), we see that increasing the number of covariates that are adjusted for increases the power. {For scenario B, where the influence of covariate $X_j$ decreases with $j$, the incremental gain in power obtained by additionally adjusting for $X_j$ (after already adjusting for $X_1,\ldots, X_{j-1}$) becomes smaller as $j$ increases.}
For scenario C (third row), we see that adjusting for more variables than $X_1$ hardly makes any difference, which is to be expected, since $X_1$ has a much stronger influence on the outcomes than the other variables in scenario C.

\begin{figure}[ht]
\centering
\includegraphics[width=0.99\textwidth]{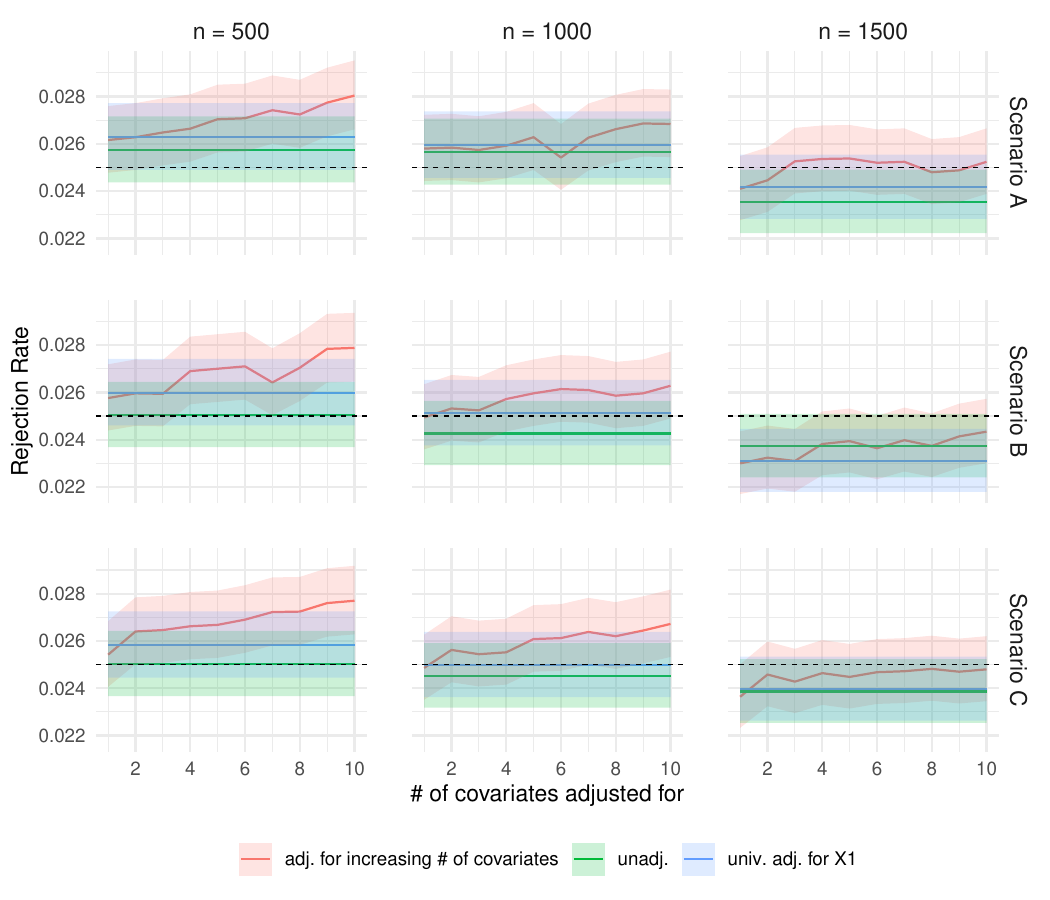}
\caption{Plots of the probability of rejecting the null hypothesis $H_0:\theta\leq 1$ at significance level $\alpha = 0.025$ with datasets of size $n \in \{500, 1000, 1500\}$ generated according to the simulation setup \eqref{eq_SimSetup} and scenarios A, B, and C for the influence of the covariates but with randomly flipped treatment assignments. The red line corresponds to adjusting for an increasing number of covariates, i.e., adjusting for $\{X_1\}$, $\{X_1, X_2\},\ldots, \{X_1, \ldots, X_{10}\}$, the green line to no adjustment and the blue line to adjusting for $X_1$ using the method from Gasparyan et al.\cite{GasparyanAdjustedWinRatio} Shaded regions indicate 95\% confidence intervals for the estimated rejection probabilities based on 50\,000 replications.}
\label{fig_SimulationH0}
\end{figure}

We also consider simulated data under the null hypothesis. That is, we generate data according to \eqref{eq_SimSetup} as before, but randomly flip the treatment assignments $A$ so that the treatment is not related to the outcome. Otherwise, we conduct exactly the same analysis as before. The results are plotted in Figure \ref{fig_SimulationH0} for sample sizes $n \in \{500, 1000, 1500\}$ and in Figure \ref{fig_SimulationH0smallN} for sample sizes $n \in \{100, 200, 300\}$. The type I error rate is close to the nominal significance level $\alpha = 0.025$ for sample sizes of $n=1500$ or if the analysis only adjusts for very few covariates. The figures show that a decrease in the sample size and an increase in the number of covariates adjusted for increases the type I error rate inflation. {We note that the relevant sample size governing the asymptotic approximation is $n$ (the number of independent observations) rather than the number of pairwise comparisons $n(n-1)$. Although $n(n-1)$ pseudo-observations enter the estimating equations, each individual participates in $n-1$ comparisons. Moreover, including many covariates in the working model relative to $n$ further destabilizes the PIM fitting and amplifies finite-sample distortions. In such settings, permutation-based inference\cite{VermeulenIncreasingThePower} may be preferred.}

\section{Discussion}
\label{sec_Discussion}
We have demonstrated how the win odds can be adjusted for baseline covariates, potentially improving the efficiency of the estimator and increasing the power of associated statistical tests. Although methodology and theory originate from Vermeulen et al.,\cite{VermeulenIncreasingThePower} the connection to the win odds used in cardiovascular outcome trials with composite endpoints (in contrast to univariate numerical outcomes) is novel. When assessing the operating characteristics of the method through a simulation study, the focus was on the specific case of time-to-event data with composite endpoints. However, the method is completely agnostic to how a win is defined, and more general endpoints and win/loss/tie rules are possible. Compared to the related literature, the method allows to perform covariate adjustment for the marginal effect (as opposed to the methods described by Dong et al.\cite{DongTheWinRatio} and Mao and Wang\cite{MaoProportionalWinFractions}) and allows for an arbitrary number of covariates (as opposed to the method described by Gasparyan et al.\cite{GasparyanAdjustedWinRatio}) including categorical variables.

The method has attractive theoretical properties. {Although it uses a working model for the conditional probabilistic index, consistency and asymptotic normality of the resulting marginal estimator do not require this model to be correctly specified.} But if it is, the estimator attains the semiparametric efficiency bound.

Through the analysis of real and synthetic data, several important practical insights emerged: i) Adjusting for highly prognostic baseline covariates can indeed increase the precision of the estimator and the power of the test, {but our empirical analyses underscored that covariate adjustment is not universally beneficial in practice. When available baseline covariates are only weakly prognostic for the outcome, the adjusted and unadjusted win odds may lead to similar inference results which is in alignment with general properties of covariate adjustment.\cite{kahan2014risks} In practice, clinical expertise and prior knowledge, including variable selection procedures performed on historical trials,\cite{wei2025current} may help identify baseline variables that are likely to be prognostic.} ii) The efficiency gains come at the cost of a slight type I error rate inflation for small sample sizes {and when a large number of covariates is adjusted for}. In such cases, alternative procedures based on permutation tests may be appropriate.\citep{VermeulenIncreasingThePower} iii) When adjusting for a single numerical covariate, the method produces results that are very close to the alternative approach of Gasparyan et al.\cite{GasparyanAdjustedWinRatio}

{A limitation of the approach, inherited from Vermeulen et al.,\cite{VermeulenIncreasingThePower} is that covariate adjustment is based on modeling the pairwise comparison probability $\Prob(Y_i \preceq Y_j | A_i, A_j, X_i, X_j)$ and subsequently averaging over the covariate distribution. In this sense, the method can be viewed as a pairwise-model-based standardization approach. This differs from classical g-computation approaches, which are based on modeling the full conditional outcome distribution $Y | A,X$ and then standardizing that model to the target population. While the pairwise formulation is natural for win statistics, it implies that the method uses a working model for the conditional probabilistic index rather than a model of the full outcome distribution. An interesting question for further research would be to investigate whether covariate adjustment methods for win statistics can be developed based directly on models for the full conditional outcome distribution $Y|A,X$. Furthermore, while misspecification of the pairwise outcome model primarily affects efficiency, it still impacts finite-sample performance, and it would also be interesting to know if one can lose asymptotic power compared to the direct estimator if the conditional model is severely misspecified. Although we did not observe settings in our experiments where adjusting clearly worsened the performance, this behavior does not seem to be completely ruled out by the theory. In addition, because the method relies on treatment-control pairwise comparisons, computation can become demanding in large trials. In this context, it is also worth noting that it is possible to conduct permutation inference without having to refit the conditional model multiple times, thereby substantially saving computational costs, see Vermeulen et al.\cite{VermeulenIncreasingThePower}}

Several further important questions could be explored in further research. One is to develop a combination of covariate adjustment with stratification. For the unadjusted win ratio and win odds, stratified analysis has been extensively studied.\cite{DongTheStratifiedWinRatio, DongTheStratifiedWinStatistics} The general idea is to only do the pairwise comparisons between the treatment and the control group within each stratum and then aggregate the results in some way. While Gasparyan et al.\cite{GasparyanAdjustedWinRatio} demonstrate how to obtain a stratified version of their adjusted win odds estimator, it is not immediately clear how to combine covariate adjustment and stratification for the method based on Vermeulen et al.\cite{VermeulenIncreasingThePower} A related question is also the connection between stratification and adjusting for a categorical variable and what exactly the difference in interpretation is. 
Cardiovascular clinical trials often include interim analysis for efficacy. Nowak et al. \cite{nowak2022group} showed that the standard group sequential theory applies to the various statistical tests for the marginal probabilistic index. In future research, the methodology presented in this manuscript could be studied in the context of group sequential designs.

Concluding, our work draws attention to the potentially useful, but yet unused connection between the literature on covariate adjustment for the MPI and the win odds. Adjusting the win odds for covariates in this way is both theoretically justified and practically feasible and can meaningfully enhance the power of clinical trial analyses where relevant baseline variables are available. We hope that this article will encourage greater awareness and adoption of these methods in future trial design and reporting.

\section*{Acknowledgements}
We are grateful to Claudio Gimpelewicz for helpful discussions about the win ratio. {Furthermore, we would like to thank two anonymous reviewers for their helpful comments and suggestions, which improved the presentation of the material.}

\section*{Conflict of interest}
Tobias M\"utze and Simon Wandel are full-time employees of Novartis Pharma AG and own stock in Novartis Pharma AG (TM, SW), Sandoz Group AG (TM, SW), and Alcon AG (SW). 
The work presented here was conducted as part of an internship project by Cyrill Scheidegger at Novartis Pharma AG.

\bibliography{sample}

@article{anker2016traditional,
  title={Traditional and new composite endpoints in heart failure clinical trials: facilitating comprehensive efficacy assessments and improving trial efficiency},
  author={Anker, Stefan D and Schroeder, Stefan and Atar, Dan and Bax, Jeroen J and Ceconi, Claudio and Cowie, Martin R and Crisp, Adam and Dominjon, Fabienne and Ford, Ian and Ghofrani, Hossein-Ardeschir and others},
  journal={European Journal of Heart Failure},
  volume={18},
  number={5},
  pages={482--489},
  year={2016},
  publisher={Wiley Online Library}
}

@article{lim2008composite,
  title={Composite outcomes in cardiovascular research: a survey of randomized trials},
  author={Lim, Eric and Brown, Adam and Helmy, Adel and Mussa, Shafi and Altman, Douglas G},
  journal={Annals of internal medicine},
  volume={149},
  number={9},
  pages={612--617},
  year={2008},
  publisher={American College of Physicians}
}

@article{bocchi2024sacubitril,
  title={Sacubitril/valsartan versus enalapril in chronic Chagas cardiomyopathy: rationale and design of the {PARACHUTE-HF} trial},
  author={Bocchi, Edimar Alcides and Echeverria, Luis E and Demacq, Caroline and de Barros e Silva, Pedro Gabriel Melo and Mazza Barbosa, Lilian and Chiang, Lu-May and Damiani, Lucas and Morillo, Carlos A and Kevorkian, Ruben and Ramires, Felix and others},
  journal={Heart Failure},
  volume={12},
  number={8},
  pages={1473--1486},
  year={2024},
  publisher={American College of Cardiology Foundation Washington DC}
}

@article{VermeulenIncreasingThePower,
author = {Vermeulen, Karel and Thas, Olivier and Vansteelandt, Stijn},
title = {Increasing the power of the Mann-Whitney test in randomized experiments through flexible covariate adjustment},
journal = {Statistics in Medicine},
volume = {34},
number = {6},
pages = {1012--1030},
year = {2015}
}

@article{ThasProbabilisticIndexModels,
author = {Thas, Olivier and Neve, Jan De and Clement, Lieven and Ottoy, Jean-Pierre},
title = {Probabilistic index models},
journal = {Journal of the Royal Statistical Society: Series B (Statistical Methodology)},
volume = {74},
number = {4},
pages = {623--671},
year = {2012}
}

@article{SongTheWinOdds,
author = {James Song and Johan Verbeeck and Bo Huang and David C. Hoaglin and Margaret Gamalo-Siebers and Yodit Seifu and Duolao Wang and Freda Cooner and Gaohong Dong},
title = {The win odds: statistical inference and regression},
journal = {Journal of Biopharmaceutical Statistics},
volume = {33},
number = {2},
pages = {140--150},
year = {2023}
}

@article{MaoProportionalWinFractions,
author = {Mao, Lu and Wang, Tuo},
title = {A class of proportional win-fractions regression models for composite outcomes},
journal = {Biometrics},
volume = {77},
number = {4},
pages = {1265--1275},
year = {2021}
}

@article{GasparyanAdjustedWinRatio,
author = {Samvel B Gasparyan and Folke Folkvaljon and Olof Bengtsson and Joan Buenconsejo and Gary G Koch},
title ={Adjusted win ratio with stratification: calculation methods and interpretation},
journal = {Statistical Methods in Medical Research},
volume = {30},
number = {2},
pages = {580--611},
year = {2021}
}

@article{WangAdjustedWinRatio,
author = {Duolao Wang and Sirui Zheng and Ying Cui and Nengjie He and Tao Chen and Bo Huang},
title = {Adjusted win ratio using the inverse probability of treatment weighting},
journal = {Journal of Biopharmaceutical Statistics},
volume = {35},
number = {1},
pages = {21--36},
year = {2025}
}

@article{PocockWinRatio,
    author = {Pocock, Stuart J. and Ariti, Cono A. and Collier, Timothy J. and Wang, Duolao},
    title = {The win ratio: a new approach to the analysis of composite endpoints in clinical trials based on clinical priorities},
    journal = {European Heart Journal},
    volume = {33},
    number = {2},
    pages = {176--182},
    year = {2011}
}

@article{PocockWinRatioInCardiologyTrials,
    author = {Pocock, Stuart J and Gregson, John and Collier, Timothy J and Ferreira, Joao Pedro and Stone, Gregg W},
    title = {The win ratio in cardiology trials: lessons learnt, new developments, and wise future use},
    journal = {European Heart Journal},
    volume = {45},
    number = {44},
    pages = {4684--4699},
    year = {2024}
}

@article{BrunnerWinOdds,
author = {Brunner, Edgar and Vandemeulebroecke, Marc and Mütze, Tobias},
title = {Win odds: an adaptation of the win ratio to include ties},
journal = {Statistics in Medicine},
volume = {40},
number = {14},
pages = {3367--3384},
year = {2021}
}

@article{DongTheWinRatio,
author = {Gaohong Dong and David C. Hoaglin and Junshan Qiu and Roland A. Matsouaka and Yu-Wei Chang and Jiuzhou Wang and Marc Vandemeulebroecke},
title = {The win ratio: on interpretation and handling of ties},
journal = {Statistics in Biopharmaceutical Research},
volume = {12},
number = {1},
pages = {99--106},
year = {2020}
}

@article{JaspersCovariateAdjustedGPC,
author = {Jaspers, Stijn and Verbeeck, Johan and Thas, Olivier},
title = {Covariate-adjusted generalized pairwise comparisons in small samples},
journal = {Statistics in Medicine},
volume = {43},
number = {21},
pages = {4027--4042},
year = {2024}
}

@article{VansteelandtGComputation,
    author = {Vansteelandt, Stijn and Keiding, Niels},
    title = {Invited commentary: G-computation–lost in translation?},
    journal = {American Journal of Epidemiology},
    volume = {173},
    number = {7},
    pages = {739--742},
    year = {2011}
}

@article{VanLanckerCovariateAdjustment,
author = {Van Lancker, Kelly and Frank Bretz and Oliver Dukes},
title ={Covariate adjustment in randomized controlled trials: general concepts and practical considerations},
journal = {Clinical Trials},
volume = {21},
number = {4},
pages = {399--411},
year = {2024}
}

@article{McMurrayAngiotensin,
author = {John J.V. McMurray  and Milton Packer  and Akshay S. Desai  and Jianjian Gong  and Martin P. Lefkowitz  and Adel R. Rizkala  and Jean L. Rouleau  and Victor C. Shi  and Scott D. Solomon  and Karl Swedberg  and Michael R. Zile },
title = {Angiotensin–Neprilysin Inhibition versus Enalapril in Heart Failure},
journal = {New England Journal of Medicine},
volume = {371},
number = {11},
pages = {993--1004},
year = {2014}
}

@article{packer2020cardiovascular,
  title={Cardiovascular and renal outcomes with empagliflozin in heart failure},
  author={Packer, Milton and Anker, Stefan D and Butler, Javed and Filippatos, Gerasimos and Pocock, Stuart J and Carson, Peter and Januzzi, James and Verma, Subodh and Tsutsui, Hiroyuki and Brueckmann, Martina and others},
  journal={New England Journal of Medicine},
  volume={383},
  number={15},
  pages={1413--1424},
  year={2020},
  publisher={Mass Medical Soc}
}

@misc{RCoreTeamR,
    title = {R: A Language and Environment for Statistical Computing},
    author = {{R Core Team}},
    organization = {R Foundation for Statistical Computing},
    address = {Vienna, Austria},
    year = {2023},
    howpublished = {https://www.R-project.org/}
  }

@misc{LumleyBiglm,
    title = {biglm: Bounded Memory Linear and Generalized Linear Models},
    author = {Thomas Lumley},
    year = {2024},
    note = {R package version 0.9-3},
    howpublished = {https://cran.r-project.org/package=biglm}
  }

@article{BebuLargeSampleInference,
    author = {Bebu, Ionut and Lachin, John M.},
    title = {Large sample inference for a win ratio analysis of a composite outcome based on prioritized components},
    journal = {Biostatistics},
    volume = {17},
    number = {1},
    pages = {178--187},
    year = {2015}
}

@article{DongTheStratifiedWinStatistics,
author = {Dong, Gaohong and Hoaglin, David C. and Huang, Bo and Cui, Ying and Wang, Duolao and Cheng, Yu and Gamalo-Siebers, Margaret},
title = {The stratified win statistics (win ratio, win odds, and net benefit)},
journal = {Pharmaceutical Statistics},
volume = {22},
number = {4},
pages = {748--756},
year = {2023}
}

@article{DongTheStratifiedWinRatio,
author = {Gaohong Dong and Junshan Qiu and Duolao Wang and Marc Vandemeulebroecke},
title = {The stratified win ratio},
journal = {Journal of Biopharmaceutical Statistics},
volume = {28},
number = {4},
pages = {778--796},
year = {2018}
}

@article{LuoAnAlternativeApproach,
author = {Luo, Xiaodong and Tian, Hong and Mohanty, Surya and Tsai, Wei Yann},
title = {An alternative approach to confidence interval estimation for the win ratio statistic},
journal = {Biometrics},
volume = {71},
number = {1},
pages = {139--145},
year = {2015}
}

@article{DongAGeneralizedAnalyticSolution,
author = {Dong, Gaohong and Li, Di and Ballerstedt, Steffen and Vandemeulebroecke, Marc},
title = {A generalized analytic solution to the win ratio to analyze a composite endpoint considering the clinical importance order among components},
journal = {Pharmaceutical Statistics},
volume = {15},
number = {5},
pages = {430--437},
year = {2016}
}

@article{OakesOnTheWinRatio,
    author = {Oakes, David},
    title = {On the win-ratio statistic in clinical trials with multiple types of event},
    journal = {Biometrika},
    volume = {103},
    number = {3},
    pages = {742--745},
    year = {2016}
}

@article{DongTheWinRatioImpactOfCensoring,
author = {Dong, Gaohong and Huang, Bo and Chang, Yu-Wei and Seifu, Yodit and Song, James and Hoaglin, David C.},
title = {The win ratio: impact of censoring and follow-up time and use with nonproportional hazards},
journal = {Pharmaceutical Statistics},
volume = {19},
number = {3},
pages = {168--177},
year = {2020}
}

@misc{FDACovariateGuideline,
  author       = {{US Food and Drug Administration (FDA)}},
  title        = {Adjusting for covariates in randomized clinical trials for drugs and biological products},
  year         = {2023},
  url          = {https://www.fda.gov/regulatory-information/search-fda-guidance-documents/adjusting-covariates-randomized-clinical-trials-drugs-and-biological-products},
  urldate      = {2025-08-15},
  note = {accessed 15 August 2025},
  howpublished = {https://www.fda.gov/regulatory-information/search-fda-guidance-documents/adjusting-covariates-randomized-clinical-trials-drugs-and-biological-products}
}

@misc{PIMPackage,
  title = {pim: Fit Probabilistic Index Models},
  author = {Joris Meys and Jan {De Neve} and Nick Sabbe and Gustavo {Guimaraes de Castro Amorim}},
  year = {2020},
  note = {R package version 2.0.2},
  howpublished = {https://cran.r-project.org/package=pim},
}

@article{VermeulenSemiparametricEstimation,
 author = {Karel Vermeulen and Jan De Neve and Gustavo Amorim and Olivier Thas and Stijn Vansteelandt},
 journal = {Statistica Sinica},
 number = {2},
 pages = {1003--1024},
 publisher = {Institute of Statistical Science, Academia Sinica},
 title = {Semiparametric estimation of probabilistic index models: efficiency and bias},
 volume = {33},
 year = {2023}
}

@article{ye2023toward,
  title={Toward better practice of covariate adjustment in analyzing randomized clinical trials},
  author={Ye, Ting and Shao, Jun and Yi, Yanyao and Zhao, Qingyuan},
  journal={Journal of the American Statistical Association},
  volume={118},
  number={544},
  pages={2370--2382},
  year={2023},
  publisher={Taylor \& Francis}
}

@article{nowak2022group,
  title={Group sequential methods for the Mann-Whitney parameter},
  author={Nowak, Claus P and M{\"u}tze, Tobias and Konietschke, Frank},
  journal={Statistical Methods in Medical Research},
  volume={31},
  number={10},
  pages={2004--2020},
  year={2022},
  publisher={SAGE Publications Sage UK: London, England}
}

@article{CaoCovariateAdjustedWinStatistic,
  title={Covariate-adjusted win statistics in randomized clinical trials with ordinal outcomes},
  author={Cao, Zhiqiang and Zuo, Scott and Baumann, Mary Ryan and Plourde, Kendra and Heagerty, Patrick and Tong, Guangyu and Li, Fan},
  journal={arXiv preprint arXiv:2508.20349},
  year={2025}
}

@article{ridker2017antiinflammatory,
  title={Antiinflammatory therapy with canakinumab for atherosclerotic disease},
  author={Ridker, Paul M and Everett, Brendan M and Thuren, Tom and MacFadyen, Jean G and Chang, William H and Ballantyne, Christie and Fonseca, Francisco and Nicolau, Jose and Koenig, Wolfgang and Anker, Stefan D and others},
  journal={New England journal of medicine},
  volume={377},
  number={12},
  pages={1119--1131},
  year={2017},
  publisher={Mass Medical Soc}
}

@article{sabatine2017evolocumab,
  title={Evolocumab and clinical outcomes in patients with cardiovascular disease},
  author={Sabatine, Marc S and Giugliano, Robert P and Keech, Anthony C and Honarpour, Narimon and Wiviott, Stephen D and Murphy, Sabina A and Kuder, Julia F and Wang, Huei and Liu, Thomas and Wasserman, Scott M and others},
  journal={New England journal of medicine},
  volume={376},
  number={18},
  pages={1713--1722},
  year={2017},
  publisher={Mass Medical Soc}
}

@article{wiviott2019dapagliflozin,
  title={Dapagliflozin and cardiovascular outcomes in type 2 diabetes},
  author={Wiviott, Stephen D and Raz, Itamar and Bonaca, Marc P and Mosenzon, Ofri and Kato, Eri T and Cahn, Avivit and Silverman, Michael G and Zelniker, Thomas A and Kuder, Julia F and Murphy, Sabina A and others},
  journal={New England Journal of Medicine},
  volume={380},
  number={4},
  pages={347--357},
  year={2019},
  publisher={Mass Medical Soc}
}

@article{GregsonHierarchical,
  title={Hierarchical composite outcomes and win ratio methods in cardiovascular trials: a review and consequent guidance},
  author={John Gregson and Dylan Taylor and Ruth Owen and Tim Collier and David J. Cohen and Stuart Pocock},
  journal={Circulation},
  year={2025},
  volume={151},
  number={22},
  pages={1606 - 1619}
}

@article{ButlerWRSeductive,
	author = {Javed Butler and Norman Stockbridge and Milton Packer},
	journal = {Circulation},
	number = {20},
	pages = {1546–1548},
	title = {Win ratio: a seductive but potentially misleading method for evaluating evidence from clinical trials},
	volume = {149},
	year = {2024}
}

@article{AjufoFallaciesOfUsingTheWR,
	author = {Ezimamaka Ajufo and Aditi Nayak and Mandeep R. Mehra},
	journal = {JACC: Basic to Translational Science},
	number = {6},
	pages = {720–727},
	title = {Fallacies of using the win ratio in cardiovascular trials: challenges and solutions},
	volume = {8},
	year = {2023}
}

@article{OConnorHFACTION,
    author = {O’Connor, Christopher M. and Whellan, David J. and Lee, Kerry L. and Keteyian, Steven J. and Cooper, Lawton S. and Ellis, Stephen J. and Leifer, Eric S. and Kraus, William E. and Kitzman, Dalane W. and Blumenthal, James A. and Rendall, David S. and Miller, Nancy Houston and Fleg, Jerome L. and Schulman, Kevin A. and McKelvie, Robert S. and Zannad, Faiez and Piña, Ileana L. and HF-ACTION Investigators, for the},
    title = {Efficacy and safety of exercise training in patients with chronic heart failure: HF-ACTION randomized controlled trial},
    journal = {JAMA},
    volume = {301},
    number = {14},
    pages = {1439-1450},
    year = {2009}
}

@misc{MaoWRPackage,
    title = {WR: Win Ratio Analysis of Composite Time-to-Event Outcomes},
    author = {Lu Mao and Tuo Wang},
    year = {2021},
    note = {R package version 1.0},
    howpublished = {https://CRAN.R-project.org/package=WR}
  }

@misc{MaoWRVignette,
  author  = {Lu Mao and Tuo Wang},
  title   = {Proportional Win-Fractions ({PW}) Regression of Composite 
             Endpoints of Death and Nonfatal Event},
  year    = {2021},
  note    = {Vignette for R package WR, version~1.0},
  howpublished     = {https://CRAN.R-project.org/web/packages/WR/vignettes/PW\_reg.html}
}

@article{kahan2014risks,
  title={The risks and rewards of covariate adjustment in randomized trials: an assessment of 12 outcomes from 8 studies},
  author={Kahan, Brennan C and Jairath, Vipul and Dor{\'e}, Caroline J and Morris, Tim P},
  journal={Trials},
  volume={15},
  number={1},
  pages={139},
  year={2014},
  publisher={Springer}
}

@article{wei2025current,
  title={Current practice on covariate adjustment and stratified analysis—based on survey results by ASA oncology estimand working group conditional and marginal effect task force},
  author={Wei, Jiawei and Mozumder, Sarwar I and Li, Liming and Xi, Dong and Xu, Jiajun and Lin, Ray and Sverdlov, Oleksandr and Chipman, Jonathan J},
  journal={BMC Medical Research Methodology},
  volume={25},
  number={1},
  pages={1--10},
  year={2025},
  publisher={Springer}
}

\appendix
\section{Additional tables and figures}\label{sec_AdditionalPlotsTables}

\begin{table}[!h]
\centering
\scriptsize
\begin{tabular}{llccc}
\toprule
Adjustment set & Statistic & unadjusted & adjusted & univ. adjusted\\
\midrule
\{Time since the index MI\} & WO & 1.038914 & 1.038692 & 1.038711\\
 & 95\% CI & (1.0034, 1.0756) & (1.0032, 1.0754) & (1.0033, 1.0754)\\
 & p-value & 0.031221 & 0.032156 & 0.032043\\
\addlinespace
\{log hsCRP at baseline\} & WO & 1.038914 & 1.040456 & 1.040520\\
 & 95\% CI & (1.0034, 1.0756) & (1.0050, 1.0771) & (1.0051, 1.0772)\\
 & p-value & 0.031221 & 0.024762 & 0.024513\\
\addlinespace
\{Dichotomized hsCRP at baseline\} & WO & 1.038914 & 1.040184 & 1.040242\\
 & 95\% CI & (1.0034, 1.0756) & (1.0048, 1.0769) & (1.0048, 1.0769)\\
 & p-value & 0.031221 & 0.025822 & 0.025586\\
\addlinespace
\{Time since the index MI, log hsCRP at baseline\} & WO & 1.038914 & 1.040249 & -\\
 & 95\% CI & (1.0034, 1.0756) & (1.0049, 1.0769) & -\\
 & p-value & 0.031221 & 0.025486 & -\\
\addlinespace
\{Time since the index MI, Dichotomized hsCRP at baseline\} & WO & 1.038914 & 1.039965 & -\\
 & 95\% CI & (1.0034, 1.0756) & (1.0046, 1.0766) & -\\
 & p-value & 0.031221 & 0.026617 & -\\
\bottomrule
\end{tabular}
\caption{Results for unadjusted, adjusted (using the methodology of Vermeulen et al.\cite{VermeulenIncreasingThePower}), and univariate adjusted \citep{GasparyanAdjustedWinRatio} win odds on the 150 mg treatment group for the synthetic CANTOS trial data. The p-values are two-sided.}
\label{tab_SynCantosResults150mg}
\end{table}

\begin{table}[!h]
\centering
\scriptsize
\begin{tabular}{llccc}
\toprule
Adjustment set & Statistic & unadjusted & adjusted & univ. adjusted\\
\midrule
\{Time since the index MI\} & WO & 1.021957 & 1.021760 & 1.021824\\
 & 95\% CI & (0.9860, 1.0592) & (0.9858, 1.0590) & (0.9859, 1.0591)\\
 & p-value & 0.234471 & 0.238681 & 0.237244\\
\addlinespace
\{log hsCRP at baseline\} & WO & 1.021957 & 1.023336 & 1.023336\\
 & 95\% CI & (0.9860, 1.0592) & (0.9874, 1.0606) & (0.9874, 1.0606)\\
 & p-value & 0.234471 & 0.205624 & 0.205534\\
\addlinespace
\{Dichotomized hsCRP at baseline\} & WO & 1.021957 & 1.022546 & 1.022559\\
 & 95\% CI & (0.9860, 1.0592) & (0.9867, 1.0598) & (0.9867, 1.0598)\\
 & p-value & 0.234471 & 0.221377 & 0.221027\\
\addlinespace
\{Time since the index MI, log hsCRP at baseline\} & WO & 1.021957 & 1.023170 & -\\
 & 95\% CI & (0.9860, 1.0592) & (0.9873, 1.0604) & -\\
 & p-value & 0.234471 & 0.208819 & -\\
\addlinespace
\{Time since the index MI, Dichotomized hsCRP at baseline\} & WO & 1.021957 & 1.022365 & -\\
 & 95\% CI & (0.9860, 1.0592) & (0.9865, 1.0596) & -\\
 & p-value & 0.234471 & 0.225064 & -\\
\bottomrule
\end{tabular}
\caption{Results for unadjusted, adjusted (using the methodology of Vermeulen et al.\cite{VermeulenIncreasingThePower}), and univariate adjusted \citep{GasparyanAdjustedWinRatio} win odds on the 50 mg treatment group for the synthetic CANTOS trial data. The p-values are two-sided.}
\label{tab_SynCantosResults50mg}
\end{table}

\begin{figure}[ht]
\centering
\includegraphics[width=0.99\textwidth]{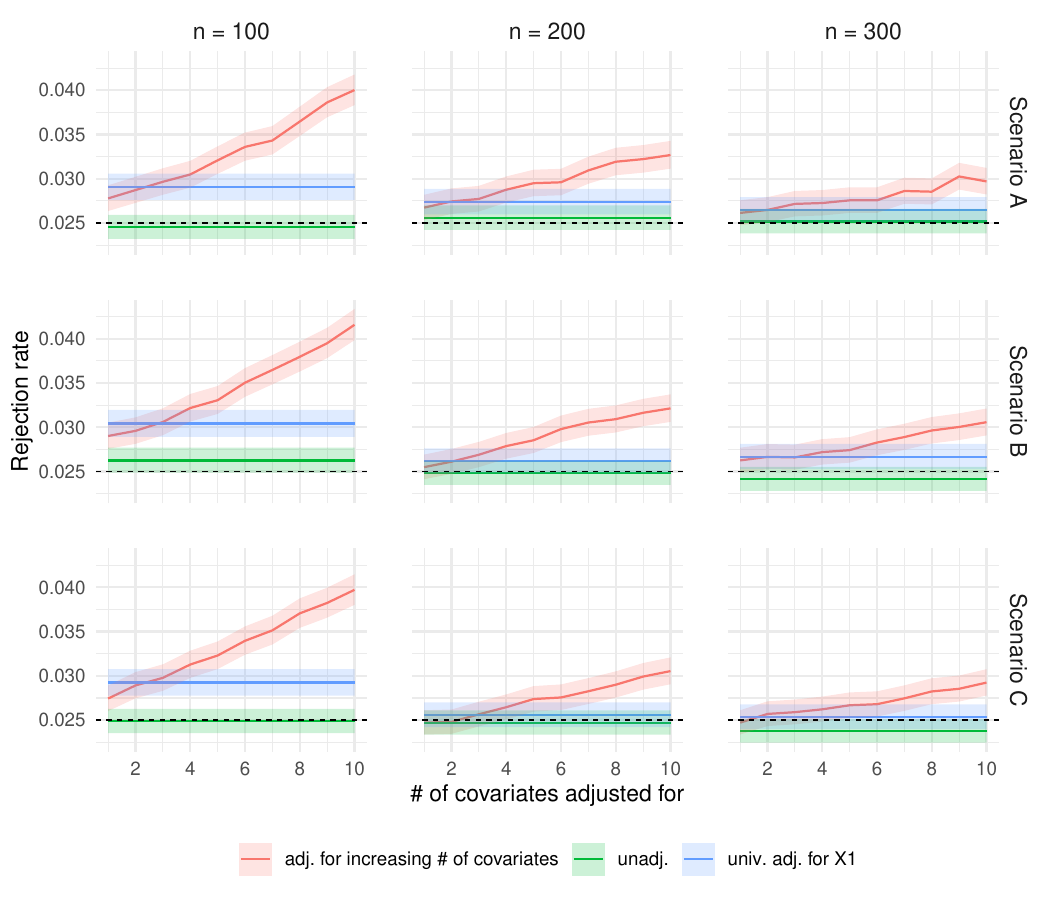}
\caption{Plots of the probability of rejecting the null hypothesis $H_0:\theta\leq 1$ at significance level $\alpha = 0.025$ with datasets of size $n \in \{100, 200, 300\}$ generated according to the simulation setup \eqref{eq_SimSetup} and scenarios A, B, and C for the influence of the covariates but with randomly flipped treatment assignments. The red line corresponds to adjusting for an increasing number of covariates, i.e., adjusting for $\{X_1\}$, $\{X_1, X_2\},\ldots, \{X_1, \ldots, X_{10}\}$, the green line to no adjustment and the blue line to adjusting for $X_1$ using the method from Gasparyan et al.\cite{GasparyanAdjustedWinRatio} Shaded regions indicate 95\% confidence intervals for the estimated rejection probabilities based on 50\,000 replications.}
\label{fig_SimulationH0smallN}
\end{figure}

\end{document}